\documentclass[conference]{IEEEtran}
\IEEEoverridecommandlockouts
\usepackage{cite}
\usepackage{amsmath,amssymb,amsfonts}
\usepackage{algorithmic}
\usepackage{graphicx}
\usepackage{textcomp}
\usepackage{xcolor}
\usepackage{booktabs} 
\def\BibTeX{{\rm B\kern-.05em{\sc i\kern-.025em b}\kern-.08em
    T\kern-.1667em\lower.7ex\hbox{E}\kern-.125emX}}

\usepackage[most]{tcolorbox}

\begin{document}

\title{Requirements Elicitation in Government Projects: \\ A Preliminary Empirical Study
}

\author{
    \IEEEauthorblockN{
    Anqi Ren\IEEEauthorrefmark{1}, 
    Lin Liu\IEEEauthorrefmark{1}, 
    Yi Wang\IEEEauthorrefmark{2}, 
    Xiao Liu\IEEEauthorrefmark{2}, 
    Hailong Wang\IEEEauthorrefmark{1},
    Kaijia Xu\IEEEauthorrefmark{1},
    Xishuo Zhang\IEEEauthorrefmark{1},
    Chetan Arora\IEEEauthorrefmark{3}
}

 \IEEEauthorblockA{\IEEEauthorrefmark{1}College of Computer Science and Technology, Inner Mongolia Normal University, Hohhot, China
   \\ 3138363067@qq.com, liulin@imnu.edu.cn, 444762194@qq.com, xishuozhang163@163.com, 9177385@qq.com,}
   \IEEEauthorblockA{\IEEEauthorrefmark{2}School of Information Technology, Deakin University, Geelong, Australia
  \\ xve@deakin.edu.au, xiao.liu@deakin.edu.au}
  \IEEEauthorblockA{\IEEEauthorrefmark{3}Faculty of Information Technology, Monash University, Melbourne, Australia
   \\ chetan.arora@monash.edu}
   
}


\thispagestyle{plain}
\pagestyle{plain}
\pagenumbering{arabic}

\maketitle

\begin{abstract}


Government development projects vary significantly from private sector initiatives in scope, stakeholder complexity, and regulatory requirements. There is a lack of empirical studies focusing on requirements engineering (RE) activities specifically for government projects. We addressed this gap by conducting a series of semi-structured interviews with 12 professional software practitioners working on government projects. These interviewees are employed by two types of companies, each serving different government departments. Our findings uncover differences in the requirements elicitation phase between government projects, particularly for data visualization aspects, and other software projects, such as stakeholders and policy requirements.
Additionally, we explore the coverage of human and social aspects in requirements elicitation, finding that culture, team dynamics, and policy implications are critical considerations. Our findings also pinpoint the main challenges encountered during the requirements elicitation phase for government projects. Our findings highlight future research work that is important to bridge the gap in RE activities for government software projects.

\end{abstract}

\begin{IEEEkeywords}
Interviews, Government Projects, Empirical Studies, Requirements Elicitation. 
\end{IEEEkeywords}

\section{Introduction}

Requirements Engineering (RE) is an important socio-technical activity involving various stakeholders and requirements in the software development life cycle~\cite{aurum2005engineering, hidellaarachchi2021effects}. Among these, the effectiveness of requirements elicitation directly impacts whether the system is successful \cite{aurum2005engineering}. Requirements elicitation commonly involves a series of approaches, such as interviews, workshops, and surveys designed to capture user needs, preferences, and constraints accurately \cite{gobov2020requirement, laplante2022requirements}. The goal is to comprehensively understand user requirements and ensure that the final product meets user expectations and business objectives \cite{pohl2010requirements}. However, the challenges associated with understanding the relationship between human aspects and regional culture remain significant during the requirements elicitation phase across various types of software projects.



Projects across government, business, and consumer domains each present unique requirements, face distinct challenges, and have specific social implications~\cite{abad2016requirements}. In this paper, we focus on government-funded data visualization projects. Hence, we use government data visualization projects and government projects interchangeably. While our findings can be generalized, that is part of our future work investigation. 
Visualization uncovers new opportunities in business and customer-focused projects, refining products and enhancing services to stimulate revenue growth~\cite{kim2021accessible}. However, meeting users' diverse data visualization preferences requires implementing multifaceted strategies to ensure inclusivity and effectiveness~\cite{qin2020making}. Furthermore, government data visualization projects can enhance the efficiency of resource allocation and service planning, enabling authorities to make informed decisions through comprehensive data analysis~\cite{mohammed2022big}. However, there are still some challenges with current government (data visualization) projects. Governments are custodians of vast repositories of sensitive information, including personal data and information critical to national security. Another challenge is the potential deficiency within government departments regarding the technical tools, expertise or financial resources to carry out complex advanced data visualization projects~\cite{lan2023intelligent}. Meanwhile, the lack of effective mechanisms for evaluating the impact of these projects and for gathering and integrating user feedback is a major challenge~\cite{graves2013visualization}.

\textcolor{black}{Government data visualization projects are data visualization solutions provided by the industrial sector to the government, aimed at addressing societal issues, such as public health crises \cite{govvisweb}.} Such projects aim to enhance internal decision-making and operational efficiency while improving public trust and interaction by transforming complex data relationships into user-friendly graphics and models~\cite{ishida2022implementation}. Data visualization faces challenges and opportunities as technology advances and societal needs change. For example, researchers focus on designing more inclusive solutions to ensure that all demographic groups, including the visually impaired and the elderly, can effectively access and understand government data visualization~\cite{ansari2022enhancing}. Government projects aim to increase public engagement by incorporating enhanced interactive designs. This includes the integration of gamification, social media connectivity, and mechanisms for real-time feedback~\cite{park2022open}. Social impact, social services, and policy are also key aspects of government data visualization projects~\cite{inastrilla2023data}. However, there is still a lack of practice for different types of government software projects during the RE phases.

As mentioned above, we need to understand the differences between government projects and other types of software projects during the requirements elicitation phase. This understanding enables us to identify the human aspects (e.g., gender, culture, team, communication, and physical issues) and social aspects (e.g., public, social impacts and policy) that are considered in RE. Human and social aspects have recently received increasing attention in RE research \cite{hidellaarachchi2021effects, Hidellaarachchi23theinf, fazzini2022characterizing, ahmad2023requirements}. Furthermore, we uncover a series of main challenges to government projects during the requirements elicitation phase.

We conducted semi-structured interviews with 12 industry practitioners, including five IT project managers, five front-end developers, and two software engineers. Six practitioners are from a medium-sized IT organization in a developed region, and others are from one medium-sized ecological data analysis company in a remote minority region. Their primary focus was on developing government data visualization projects. All practitioners had over three years of industry experience.

Our findings revealed that government data visualization projects differ from other software projects. In government projects, the intersection of technology, policy, and social impact plays a critical role in shaping project outcomes and stakeholder satisfaction. Unlike customer-focused projects, where requirements are directly sourced from end-users, government projects often involve a complex array of stakeholders, including government employees, policymakers, and the public. This diversity necessitates a more inclusive and multifaceted approach to requirements elicitation, where considerations of culture, team dynamics, and policy implications come to the forefront. Our findings indicated that the requirements elicitation phase incorporates a variety of human and social aspects considerations, with culture, team, and policy being key considerations. Our findings also identified several challenges, including a lack of professional knowledge, technical issues, social issues, ambiguity, vagueness, and changes in requirements. Finally, we discuss the findings and propose research for future work.

   

The remainder of this paper is structured as follows: Section \ref{rw} reviews the related work. Section \ref{reM} introduces the methods used in our study. Section \ref{results} presents the findings. Section \ref{Dis} presents the discussion. Section \ref{threats} presents threats to the validity of this study. Finally, Section \ref{conc} concludes the paper and proposes future research work.

\section{Related Work}\label{rw}

Requirements Engineering (RE) places a strong emphasis on capturing and analyzing user requirements, which significantly influences project outcomes. Insights from various disciplines indicate that data visualization plays a crucial role in the understanding of data, information, and knowledge \cite{abad2016requirements,arora2310advancing}. In recent years, some studies have investigated the role of data visualization within the software engineering life cycle, focusing on areas such as design process \cite{horcas2022variability, horcas2023empirical} and product lines \cite{walny2019data, hinterreiter2020visualizing}. Zhang et al.\cite{zhang2023personagen} developed an automatic generating persona tool that uses the GPT-4 model and knowledge graph visualization to generate persona templates from well-processed user feedback, facilitating requirement analysis. This tool employs basic visualization graphs, such as bar charts, to display the categorization of different user attributes. Additionally, there are mature visualization applications and tools. For example, Bulej et al. \cite{bulej2020ivis} developed the IVIS platform, which features a web-based interface enabling the definition of complex visualizations and data processing tasks, as well as data exploration. Distinct from existing open-source and commercial solutions, IVIS adopts a unique model emphasizing flexibility. Overall, although these studies have highlighted the connection between RE and visualization, the integration of visualization techniques within in requirements elicitation phase has not been extensively explored \cite{Abad16}. If user requirements are not fully understood during the requirements elicitation phase, it may lead developers to misunderstand these requirements, ultimately resulting in user dissatisfaction with the software product. Meanwhile, employing visualization techniques in requirements elicitation phase can enhance the capture of user requirements, thereby benefiting subsequent product design and deployment \cite{walny2019data}.







Government data visualization projects are aimed at enhancing transparency, efficiency and effectiveness of decision-making \cite{ansari2022enhancing}. These projects enable both government employees and the public to more effectively understand and analyze information by transforming complex data in a visually friendly manner~\cite{gottfried2021mining}. In recent years, although open government data (OGD) have become increasingly available, their full potential has yet to be reached due to their limited utilization. Incorporating visualization into open government data can render the data more engaging, useful, and accessible to users \cite{ansari2022enhancing}. Matheus et al.\cite{matheus2020data} proposed an approach to enhance the efficiency of requirements elicitation by understanding and supporting the design of dashboards. These dashboards are designed to visualize data pertinent to requirements elicitation, enabling users to monitor ongoing activities and initiate necessary actions.
Additionally, policymakers can use dashboards to support their decision-making and policy formulation processes or to facilitate communication and interaction with the public. Chokki et al. \cite{chokki2022engaging} compiled a comprehensive list of 16 dashboard design principles for the context of OGD, which informed the development of the Namur (Belgium) Budget Dashboard (NBDash). Through experimentation with NBDash, including the selection of meaningful metrics and the employment of suitable visualizations within a coherent presentation framework, these principles have been validated as critical to the success of dashboards. However, dashboards can encounter challenges, including low data quality, insufficient understanding of the data, and the potential for incorrect analysis. However, to enhance the efficiency of RE and increase user satisfaction, there is a need for visualization tools that can accurately capture requirements and analyze~\cite{li2021blockchain}. Additionally,  while many studies have focused on using visualization technologies to facilitate RE activities, there remains an important gap in empirical evidence concerning the influence of RE activities on the design and development of data visualization systems. 

In summary, there are many challenges to be addressed in government data visualization projects. However, no study (to the best of our knowledge) has covered the practitioners' perspective in eliciting user requirements for government data visualization projects. Our study reveals the differences between government data visualization projects and other software projects, considerations of human and social aspects, and a series of main challenges during the requirements elicitation.


\section{Research Method}\label{reM}

Our study aims to understand the practitioners' perspectives on requirements elicitation in government projects, focusing on data visualization projects in China. We base our study on semi-structured interviews to collect different types of self-report data. The Human Ethics Advisory Group at our university approved our study. Our interview material is publicly available in both Chinese and English~\cite{survey}. 

\subsection{Research Questions}

Our study was to investigate the differences between government data visualization projects and other software projects, consider the human and social aspects, and identify the main challenges in the requirements elicitation phase. The three research questions (RQs) that guide our study are: 


\textbf{\textit{RQ1.}} What are the differences in the requirements elicitation phase between government data visualization projects and other software projects?

\textbf{\textit{RQ2.}} Which human and social aspects are considered during the requirements elicitation phase for government data visualization projects?

\textbf{\textit{RQ3.}} What are the main challenges in requirements elicitation for government data visualization projects?

\subsection{Participants Recruitment}

For this study, we recruited all interviewees through our professional networks, specifically by advertising our recruitment via companies we are currently collaborating with. We conducted interviews with 12 professionals (P1-P12), representing variety of job roles within one medium big data visualization company and one medium IT company. These two companies served the government departments (environmental protection department, and meteorological department, respectively) in Inner Mongolia, China. \textcolor{black}{Specifically, each government department involves participation from technical divisions and other employees (holding non-fixed job positions and possessing a basic IT knowledge).} Among them, the big data visualization company is located in a remote minority region, while the IT company is located in a developed region. \textcolor{black}{All interviews have more than three years of experience working in industry, and nigh practitioners (P1-P7, P10 \& P11) always participated requirement tasks in government data visualization projects and three practitioners rarely participated (P8, P9 \& P12).} \textcolor{black}{Two practitioners lack prior experience in business and user software projects (P3 \& P6). All interviewees have experience in data visualization projects, among eight practitioners have experience in participating other types of software applications (e.g., video game, extended reality based interactive system, and customer relationship management system). }  Each interviewee received an incentive of 80 RMB e-gift cards. Table \ref{tab:interview} shows the demographic information of each participant.

\begin{table}
    \centering
  \caption{Overview of the interview participants \textit{(N = 12).}}
    
    \label{tab:interview}
    \begin{tabular}{clp{3cm}c}
 \toprule

        ID & Occupation & Company & Exp. Years \\        \midrule 
      P1 &IT Project Manager &Ecological data analytics company  &12   \\
      P2 &IT Project Manager & Ecological data analytics company  &5   \\
      P3   &IT Project Manager  & Ecological data analytics company &3  \\
      P4   &Front-end Developer  & IT company  &5  \\
      P5   &Front-end Developer  &IT company &8  \\
      P6   &Front-end Developer  & IT company &3  \\
      P7   &Front-end Developer  & IT company  &3  \\
      P8   &Software Engineer  & Ecological data analytics company &10  \\
      P9   &Software Engineer  & Ecological data analytics company  &8  \\
      P10  &IT Project Manager  & Ecological data analytics company &6  \\
      P11   &IT Project Manager  & IT company &4  \\
      P12   &Front-end Developer   & IT company  &5  \\
 \bottomrule
    \end{tabular}
\vspace*{-1.5em}
 \end{table}   

\subsection{Semi-structured Interview Design}

The first author conducted a series of online interviews with eight practitioners involved in government projects, ensuring that at least two co-authors were present in every semi-structured interview. Meanwhile, two co-authors (second and third authors) also interviewed four practitioners. Each interview lasted between 30 to 45 minutes. Before conducting the interview, we designed an interview guideline to ensure that all questions were covered within a maximum of 60 minutes. To ensure the quality of the interview data, we also iteratively refined our interview guidelines after the second and seventh interviews. The interview guideline had three parts:

In the first part, we first introduced the objectives and goals of our study. We then asked demographic questions about the interviewees' experience with RE activities for government projects, emphasizing requirements elicitation. We also asked about their years of experience and job roles with their respective companies.

In the second part, we asked stakeholders involved in government data visualization projects about the methods and approaches used to elicit user requirements and identity user requirements. We also asked interviewees to provide recent examples of government data visualization projects and discussed the differences in eliciting user requirements between government data visualization projects and other types of visualization projects, such as business or customer projects.

In the third part, we asked the interviewees to discuss whether to consider human and social aspects in eliciting user requirements for government data visualization projects. Specifically, we introduced the concept of human and social aspects in the processes of eliciting user requirements, as followed by Hidellaarachchi et al. \cite{hidellaarachchi2021effects, Hidellaarachchi23theinf}. Considering the cultural diversity and inclusivity in China, we specifically asked interviewees to discuss the influence of cultural aspects in eliciting user requirements for government data visualization projects. Additionally, we asked interviewees to discuss about the consideration of accessibility requirements when eliciting user requirements from disabled users for government data visualization projects.

In the fourth part, we asked the main challenges faced during the requirements elicitation phase for current government data visualization projects. Additionally, we asked the interviewees general questions to the interviewees, such as \textit{``Are there any additional suggestions you would like to offer regarding the elicitation of user requirements for government data visualization projects?''} and \textit{``Do you have any questions for us?''} At the end of each interview, we thanked the interviewees for their participation.


\subsection{Data Analysis}

To analyze the interview data, we used the iFLYTEK professional automated transcription service to convert the online interview recordings into transcripts and translated the Chinese transcript into English~\cite{ifly23}. The first author and two co-authors performed an accuracy check on the auto-transcriptions. They are proficient in both English and Chinese. Then, we conducted a thematic analysis to code the entire transcripts using MAXQDA, following recommended practices~\cite{maxqad24, cruzes2011recommended, gizzi2021practice}. The first author and two co-authors first reviewed all transcripts, created memos on topics relevant to research questions, and assigned labels to transcripts, referred to as codes. To ensure the quality of the codes, three co-authors checked the initial codes created and provided suggestions for refinement. After incorporating these suggestions, we generated a total of 134 codes. The authors then merged the codes with the same words or meanings. After merging the codes code, we have a total of 113 unique codes and three main themes. \textcolor{black}{Among the identified themes, most are directly related to the primary themes and their sub-themes. However, we also found some primary themes and sub-themes with latent errors. Three participants reviewed the context and consequently removed one latent primary theme along with five sub-themes, and five sub-themes in primary themes.} The primary themes include \textit{Differences}, \textit{Human and Social Aspects}, and \textit{Main Challenges in Government Data Visualization Projects}. The themes and findings are discussed in Section~\ref{results}.


\section{Findings}\label{results}

\tcbset{
    frame code={}
    center title,
    left=19pt,
    right=0pt,
    top=-6pt,
    bottom=-6pt,
    opacityback=0,
    width=234pt,
    enlarge left by=0mm,
    arc=6pt,outer arc=0pt,
    }

We present our findings from analyzing the data collected from the interviews.

\subsection{RQ1 Results - Differences}\label{RQ1}

In this section, we report the differences in eliciting user requirements between government data visualization projects and other types of software projects (RQ1).

\textbf{\textit{Stakeholders. }}Our interviewees indicated that stakeholder participation was a different aspect between government data visualization projects and other types of software projects. Most interviewees emphasized that government employees play a crucial role in defining software requirements (P1-P9, P11, P12). Some interviewees highlighted the significant contribution of the public (users) in the software requirements definition process (P2, P4, P7, P11). However, they may not be directly involved in the early requirements elicitation activities for government data visualization projects. P2, P7 and P12 also explained that government data visualization projects may be designed to serve both external users and businesses, as well as to serve intra-governmental initiatives. P4 also reported a relative lack of control in such projects, e.g., in a current government data visualization project is collaborating with various organizations and departments, including \textit{``government departments (protection and ecological departments)''}, \textit{``universities''}, and \textit{``business companies (IT and ecological data analytics companies)''}, which the funding department dictates. P1 and P12 also considered that:

\begin{tcolorbox}
\textit{``...it`s up to them (head of government departments) to figure out what's needs (regarding data display), because they're the ones who know what visitors (users) or the government employees are looking to find out.'' (P1)}
\end{tcolorbox}

\begin{tcolorbox}
\textit{``...government data visualization projects have a diverse group, including minority nation groups, stakeholders, and organizations with complex preferences and needs.'' (P12)}
\end{tcolorbox}

However, P2 and P4 highlighted that eliciting requirements from government employees present challenges due to their potential lack of experience and professional knowledge in relevant technology (see Section~\ref{challenges} for details).

\textbf{\textit{Policy Requirements. }} Many interviewees indicated that the policy requirements were different between government data visualization projects and other types of software projects (P1-P6, P8-P11). P8 indicated that government data visualization projects have more stringent requirements for policy and legal. P9 explained that:

\begin{tcolorbox}
\textit{``When we're figuring out needs, we have to think about all the different people making decisions and what the policies say.'' (P9)}
\end{tcolorbox}

However, some interviewees mentioned that policy changes are an important factor influencing requirements changes during the requirements elicitation phase (P8, P9, P10, P11). P9 highlighted that policy priorities are one of the key factors in the requirements elicitation phase. P10 indicated that:

\begin{tcolorbox}
\textit{``Requirements are key for government data visualization projects because they have to stick to strict policies and laws.'' (P10)}
\end{tcolorbox}

Additionally, P4 reported that government data visualization projects require higher transparency, in stark contrast to business or customer projects, which prioritize confidentiality during the requirements engineering phase.

\textbf{\textit{Time and Quality. }}Some interviewees indicated that time and quality differ in the requirements elicitation phase between government data visualization projects and other software projects (PP6, P8, P9, P11, P12). Also, there are waves of initiatives, e.g., the recent wave of digital transformation -- which are difficult to implement in reality. P6 noted that government data visualization projects are less agile and require more time during the requirements elicitation phase, for example:

\begin{tcolorbox}
\textit{``I might have to do multiple rounds to confirm the requirements. This process takes longer, often need a month or two month to wrap up.'' (P6)}
\end{tcolorbox}

P5 explained that government employees have primary jobs and may be unable to focus on digital transformation initiatives and the corresponding software projects. Conversely, a few interviewees indicated that some government data visualization projects may require a higher quality of user requirements during the requirements elicitation phase (P1, P5, P10). To explain this point, P5 provided an example that:

\begin{tcolorbox}
\textit{``For (anonymous) projects, the overall timeline is tighter. So, we need to be extra careful and detailed during the requirements gathering phase to make sure we hit the mark with what users expect.'' (P5)}
\end{tcolorbox}


\subsection{RQ2 Results - Human and Social Aspects}

This section presents the human and social aspects involved in understanding user requirements during the requirements elicitation phase of government data visualization projects that we identified in our interview study (RQ2).

\begin{figure}[!t]
  \centering
  \includegraphics[width=0.96\linewidth]{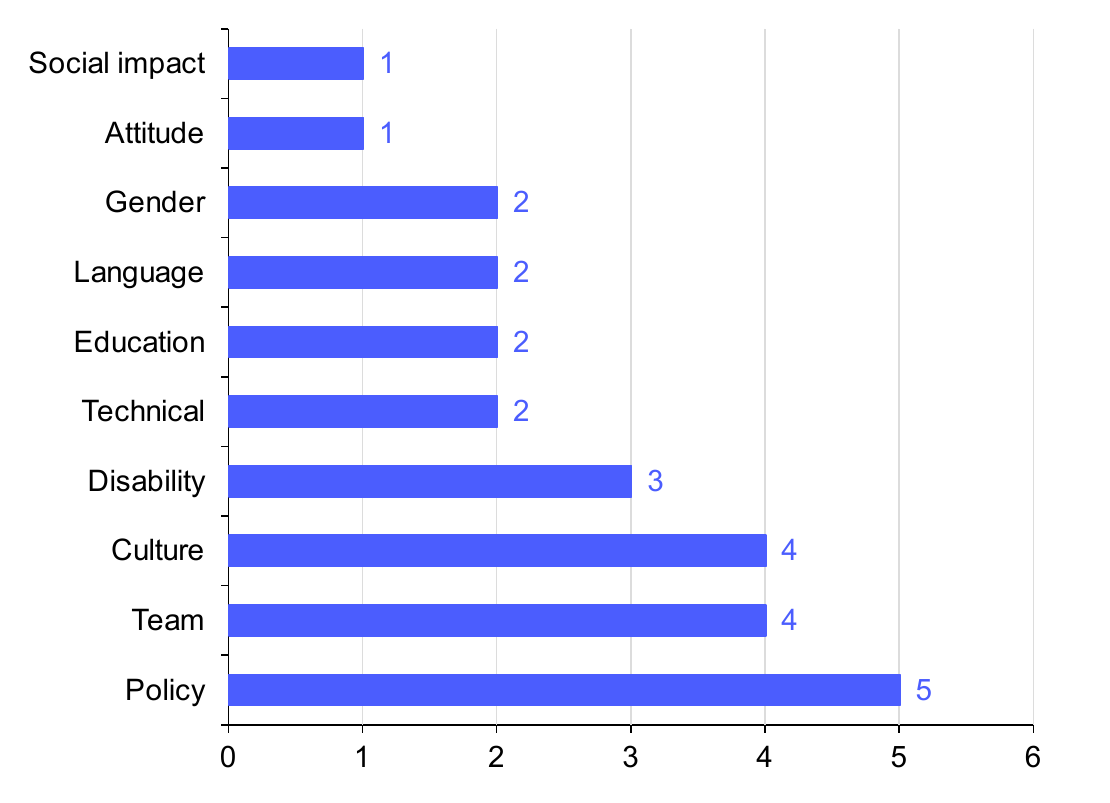}
  \vspace*{-1em}
  \caption{Key human and social aspects mentioned by interviewees. }
   \label{fig:human}
   \vspace*{-1em}
\end{figure}

Many interviewees indicated that government data visualization projects should consider various human and social aspects during the requirements elicitation phase (P1-P3, P5-P9, P12). Figure \ref{fig:human} shows the human and social aspects mentioned by interviewees during the requirements elicitation phase for government data visualization projects.

\textbf{\textit{Individual Human Aspects. }}Some interviewees mentioned that individual human aspects are crucial considerations in the requirements elicitation phase for government data visualization projects (P1, P3, P5, P8). Some interviewees indicated that \textit{Culture}, \textit{Education}, \textit{Language}, \textit{Gender}, \textit{Disability}, and \textit{Attitude} were considered during the requirement elicitation phase (P3, P5, P6, P8). P2, P3 and P5 respectively stated that:

\begin{tcolorbox}
\textit{``Our government project serves minority ethnic areas. We need to consider local languages, cultures, and educational backgrounds.'' (P2)}
\end{tcolorbox}

\vspace{-0.1cm}

\begin{tcolorbox}
\textit{``We've got to think about cultural differences, gender equality, education levels, what people with disabilities need, and age differences. These factors can change how users feel about and accept different ways of showing, and interactive.'' (P3)}
\end{tcolorbox}

\vspace{-0.1cm}

\begin{tcolorbox}
\textit{``People from various regions and cultural backgrounds may have different perspectives on how to interpret and use data. Considering cultural factors in requirements can reduce cultural sensitivity and conflicts.'' (P5)}
\end{tcolorbox}

P3 also stated the necessity of collaboration with local universities and government departments due to a lack of relevant knowledge among project stakeholders. However, there was a series of collaborative challenges across different departments and organizations (see \ref{challenges} for details). Furthermore, although five interviewees mentioned the importance of considering the accessibility requirements of users with disabilities (P1, P2, P3, P5, P11), only one interviewee indicated that accessibility requirements were considered in such projects (P5). P5 indicated that government employees only proposed simple accessibility requirements, such as text prompts. P6 and P11 stated that:

\begin{tcolorbox}
\textit{``Accessibility wasn't factored into the many government projects. Other visualization projects (business or customer projects) have thought about users like the elderly or kids, focusing on making pages simple and user-friendly.'' (P11)}
\end{tcolorbox}

\vspace{-0.1cm}

\begin{tcolorbox}
\textit{``The needs of people with disabilities might impact how accurately data is presented.'' (P6)}
\end{tcolorbox}

Some interviewees considered pressing time constraints and limited funding as key factors leading to the neglect of accessibility requirements (P4, P7, P9). Additionally, P6 noted that stakeholders' attitudes influence requirements elicitation phase. P5 also mentioned that government employees may not be able to focus on types of software projects.

\textbf{\textit{Team Human Aspects. }}Some interviewees mentioned that team collaboration is a key aspect in the requirements elicitation phase for government data visualization projects (P3, P6, P8, P9). P3 emphasized the need for multiple departmental collaboration, involving IT project managers, designers, developers, data analysts, and government employees in the requirements elicitation phase. Additionally, P3 stated that government departments may prioritize to seek collaboration with local IT companies. However, due to regional variations in technology and knowledge, some government data visualization projects aim to collaborate with IT companies or universities from other regions. Meanwhile, P6,  and P8 emphasized that regional differences, including skills, experience, culture, and knowledge, can significantly affect the quality of requirements during the requirements elicitation phase. Additionally, P9 indicated that:

\begin{tcolorbox}
\textit{``Requirements activities need to team up tightly with government departments, stakeholders, and people in the community to make sure the project keeps meeting what users need and makes a good impact on society.'' (P9)}
\end{tcolorbox}

\textbf{\textit{Social Aspects. }} Some interviewees mentioned that policy aspect are a key consideration during the requirements elicitation phase (P3, P4, P5, P7, P9, P10). They indicated that government data visualization projects not only focus on stakeholder requirements but also comply policy and legal requirements. P3 also stated that:

\begin{tcolorbox}
\textit{``This means we have to cover requirements like protecting data privacy, keeping information secure, and making sure everything is open and transparent.'' (P3)}
\end{tcolorbox}

Additionally, P9 highlighted the importance of considering the social impacts and strategic goals of government departments during the requirements elicitation phase.

\subsection{RQ3 Results - Challenges}\label{challenges}

In this section, we present the main challenges of eliciting user requirements for government data visualization projects that we identified in our interview study (RQ3).

\textbf{\textit{Ambiguity, Change, and Vagueness in Requirements. }}Some interviewees indicated that ambiguity in requirements represents an important challenge during the requirements elicitation phase (P3, P4, P6, P7). P4 noted that coordinating different perspectives from various departments and stakeholders and reaching a consensus are a challenge. P4 also reported an example that:

\begin{tcolorbox}
\textit{``...(anonymous) project, when collecting requirements from the environmental protection department, and meteorological department, we found there were quite a few ambiguities in their needs for data presentation and understanding.'' (P4)}
\end{tcolorbox}

P3 indicated that ambiguous requirement descriptions provided by government employees can result in ambiguities during the requirements elicitation phase. P3 further suggested that development teams can mitigate this issue by refining requirements through iterate clarification and definition, achieved by actively engaging in questions. Conversely, P6 indicated that government employees are unable to accurately describe their requirements. As a solution, they suggested continuously setting different scenario-based questions to encourage employees to specify their needs more clearly. Additionally, P9 reported frequently encountering vagueness and ambiguous requirements descriptions.

Some interviewees indicated that government data visualization projects often face vagueness and changing requirements (P3, P5, P6, P8, P9). P8 indicated that policy, strategy, and user feedback are the main factors in requirements changes.

\textbf{\textit{Social Issues. }}Some interviewees indicated that public interest, social, and cultural considerations are major challenges during the requirements elicitation phase (P2, P3, P5). Specifically, P2 indicated that:

\begin{tcolorbox}
\textit{``Government data visualization projects usually deal with the public's interest and involvement, so figuring out how to really include what people need and think is a big challenge.'' (P2)}
\end{tcolorbox}

P3 emphasized that characteristics of diverse user groups, including culture, gender, education, and people with disabilities, significantly vary across societal contexts. P5 indicated that:

\begin{tcolorbox}
\textit{``The user base for government services is really wide, covering various socio-economic statuses and cultural differences. Understanding and catering to these diverse needs is a challenge.'' (P5)}
\end{tcolorbox}

\textbf{\textit{Technical Issues. }}Some interviewees noted that technical issues are a key challenge (P4, P8, P9, P12). P4 indicated that adoption of unique technical standards and infrastructure by government departments can significantly influence the development and integration of visualization systems. P8 stated that:

\begin{tcolorbox}
\textit{``Previous government projects have used outdated tech or framework, which might not be suitable for new government visualization projects'' (P8)}
\end{tcolorbox}

Furthermore, P9 mentioned that considering the needs of user with disabilities faces many challenges.

\begin{tcolorbox}
\textit{``This involves balancing the needs of disabled users and normal users, tackling the complexities and data density, and the constraints of technical implementation.'' (P9)}
\end{tcolorbox}

Additionally, P8 indicated that data security and privacy are important factors for government data visualization projects, thus requiring stringent compliance and confidentiality measures for both users and data.

\textbf{\textit{Insufficient Professional Knowledge. }}Some interviewees mentioned insufficient professional knowledge as a key challenge (P1, P2, P5, P9). Specifically, P1 highlighted the issue of government employees' lack of professional knowledge. 

\begin{tcolorbox}
\textit{``Actually, one of the major issues is that what leaders want to showcase isn't always supported by available data in the system, or there's simply no way to obtain the right data.'' (P1)}
\end{tcolorbox}

P2 noted that stakeholders, including government employees and project managers, struggle with understanding newer concepts, such as data warehouses, virtualization analysis, and relationships among data, information, and knowledge.

\begin{tcolorbox}
\textit{``The meanings of these concepts and terms can vary and be unclear because of different specializations and backgrounds, needing extra time and work to make them clear.'' (P2)}
\end{tcolorbox}

P9 emphasized the necessity for teams to have professional knowledge, the ability to analyze and evaluate data, and a comprehensive understanding of visualization technologies. Furthermore, P5 point out that government data visualization projects often face the challenge of lacking systematic user research and data analysis, which can result in a deficient grasp of user needs.

\section{Discussion}\label{Dis}

In this section, we reflect on the findings of the research questions and discuss the differences between government data visualization projects and other types of software projects, considering the human and social aspects and challenges.


The study findings confirm that stakeholder differences are an important factor in distinguishing government data visualization projects from other software projects. Specifically, \textbf{the definition of requirements in government data visualization projects is often oligarchic in nature, i.e., primarily influenced by government department leaders.} 

Government data visualization projects commonly involve a range of stakeholders, including government employees, the general public, academic researchers and students, as well as various organizations and companies. However, \textbf{government employees often lack professional knowledge and are not up to date on the latest technologies}, leading to quality defects (e.g., incompleteness and vagueness) in requirements in the requirements elicitation phase. 
Another related theme in our analysis that we encountered is the trend of digital transformation in such projects. On one hand, it enables innovative approaches to citizen engagement, leveraging data visualization as a tool for transparency and informed decision-making. On the other hand, it underscores the critical challenges of safeguarding sensitive information, bridging the digital divide, and ensuring that technological solutions are accessible and equitable, with the issue that most key stakeholders are usually not updated on the latest technologies.

Next, government projects tend to uniquely prioritize policy considerations in requirements. This underscores the critical role of government policies in driving changes to project requirements. Our interviews also revealed that, \textbf{compared to non-government projects, government initiatives demand significantly more effort in the requirements elicitation phase}. This increased effort can be attributed to various constraints, including limited time, policy restrictions, and budgetary limitations. The success of some projects may be hindered by challenges in stakeholder collaboration, such as insufficient professional knowledge and the limited time allocated by governmental stakeholders to such projects. As a result, requirements engineers require additional time to accurately capture and comprehend user requirements.

In the requirements elicitation phase of \textbf{government data visualization projects, it's essential to take into account a broad spectrum of human and social factors due to the potential impact of such projects on a large cohort of citizens}. Cultural considerations are often intertwined with other personal human factors, such as language, gender, and educational background, influencing the project requirements significantly~\cite{hidellaarachchi2021effects, Hidellaarachchi23theinf}. Hence, it is crucial for practitioners to understand the local cultural background and address the requirements of the local community, including remote minority groups. We further observe (in our context) that regional differences have a profound effect on the success and quality of these projects. Teams located in remote minority regions may lack the skills and expertise necessary for the development of such projects and, hence, are not the top choice for leading - despite being the main beneficiaries of such projects. Conversely, teams from non-minority regions lack an understanding of the local cultural background and policies. Therefore, bridging regional disparities through effective team collaboration becomes critically important during the requirements elicitation phase of such projects.

\section{Threats to Validity}\label{threats}

\textbf{\textit{Construct Validity}} is a crucial aspect of validation in qualitative analysis. Two co-authors had sound research experience in RE; one primarily focused on human-centric RE, while the other had solid industrial experience in data visualization development. They used their expertise to assist in creating an accurate interview guide and research objectives. To reduce interview bias, they also conducted interviews with four interviewees. Additionally, the interview guide was improved in terms of relevance and clarity through multiple iterations. 

\textbf{\textit{Internal Validity.}} We had to make an opportunistic selection of interviewees based on the recommendations by company leaders' recommendations, but we required interviewees to have at least three years of work experience. To reduce selection bias, interviewees had no prior contact with researchers before the interviews. 

\textbf{\textit{External Validity.}} We interviewed 12 interviewees from China. While the overall cultural difference may be relatively minor, but regional cultural differences could still be significant. Although our study is focused on investigating the elicitation of user requirements for government data visualization projects in remote minority areas, these findings can provide valuable insights for researchers and practitioners in human-centric requirements engineering, especially those involved in government software projects.

\textbf{\textit{Reliability.}} To minimize these risks, we had multiple researchers ensure the accuracy of the transcript and data reliability. Three authors discussed coding results to ensure data consistency over multiple rounds. Nonetheless, some degree of subjectivity may still exist. Additionally, as our focus is on practitioners' insights into government projects, specifically regarding government data visualization projects during the requirement elicitation phase, our study lacks a comparative analysis of the practical distinctions between government projects and other types of software projects across various RE phases. We leave the systematic comparison to future work.

\section{Conclusion and Future Work}\label{conc}

This paper investigates the difference between government projects and other software projects and explores the consideration of human and social aspects as well as the challenges encountered during the requirements elicitation phase. Our findings imply that there are some different aspects of the requirements elicitation phase, particularly concerning stakeholders and policy requirements. Culture, team and policy are important considerations during the requirements elicitation phase. Our findings also confirm a range of challenges. For example, stakeholders, including government employees, may lack professional knowledge, resulting in ambiguity and vagueness in requirements and there can be a general lack of agility in such projects. Additionally, policy changes are one of the primary factors driving changes in requirements. As ongoing work, we proposed a series of future research work:

\begin{itemize}
  \item We plan to extend our survey to gain a comprehensive understanding of RE activities across a range of government projects, including data visualization and extended reality (XR) and other emerging technologies, particularly in different regions, in particular the remote minority regions in China. This will enable us to capture a diverse set of practitioner perspectives and understand the practices and challenges of different types of government projects.
  
  
  \item We plan to conduct a comprehensive investigation of how human factors, regional cultures, and considerations for individuals with disabilities influence RE activities.

  \item \textcolor{black}{We plan to conduct a comprehensive exploration of the differences between government projects and other software projects during the RE phases and the challenges faced by stakeholders working on government software projects. This will facilitate both researchers and industry practitioners in understanding the opportunities and challenges present in government projects in RE phases.}

\end{itemize}

Finally, our goal is to bridge the gap in human-centric requirements engineering for government software projects. 


\bibliographystyle{IEEEtran}

\bibliography{ref}

\end{document}